# Local distortions and dynamics in hydrated Y-doped BaZrO$_3$

*Amangeldi Torayev [1], Luke Sperrin [1], Maria A. Gomez [2], John A. Kattirtzi [1], Céline Merlet\* [1,3,4], Clare P. Grey\* [1]*

[1] Department of Chemistry, University of Cambridge, Lensfield Road, Cambridge CB2 1EW, United Kingdom

[2] Department of Chemistry, Mount Holyoke College, South Hadley, MA 01075, United States

[3] CIRIMAT, Université de Toulouse, CNRS, Université Toulouse 3 - Paul Sabatier, 118 Route de Narbonne, 31062 Toulouse cedex 9 - France

[4] Réseau sur le Stockage Electrochimique de l'Energie (RS2E), FR CNRS 3459, HUB de l'Energie, Rue Baudelocque, 80039 Amiens, France

**Corresponding Authors:** merlet@chimie.ups-tlse.fr, cpg27@cam.ac.uk

**ABSTRACT**

Y-doped BaZrO$_3$ is a promising proton conductor for intermediate temperature solid oxide fuel cells. In this work, a combination of static DFT calculations and DFT based molecular dynamics (DFT-MD) was used to study proton conduction in such a material. Geometry optimisations of 100 structures with a 12.5% dopant concentration allowed us to identify a clear correlation between the bending of the metal-oxygen-metal angle and the energies of the simulated cells. Depending on the type of bending, two configurations, designated as inwards bending and outwards bending, were defined. The results demonstrate that a larger bending decreases the energy and that the lowest energies are observed for structures combining inwards bending with protons being close to the dopant atoms. These lowest energy structures are the ones with the strongest hydrogen bonds. DFT-MD simulations in cells with different yttrium distributions provide complementary microscopic information on proton diffusion as they capture the dynamic distortions of the lattice caused by thermal motion. A careful analysis of the proton jumps between different environments confirmed that the inwards and outwards bending states are relevant for the understanding of proton diffusion. Indeed, intra-octahedral jumps were shown to only occur starting from an outwards configuration while the inwards configuration seems to favor rotations around the oxygen. On average, in the DFT-MD simulations, the hydrogen bond lengths are shorter for the outwards configuration which facilitates the intra-octahedral jumps. Diffusion coefficients and activation energies were also determined and compared to previous theoretical and experimental data showing a good agreement with previous data corresponding to local proton motion.

**TOC GRAPHICS**

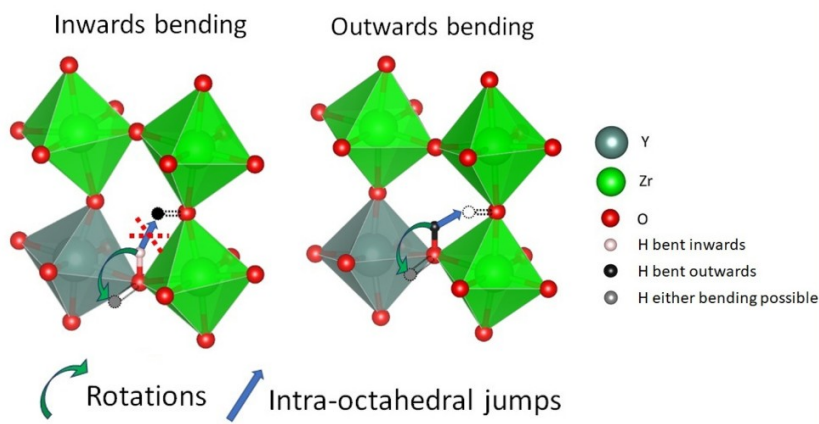

**1. Introduction**

Perovskites have emerged as attractive electrolytes for intermediate-temperature Solid Oxide Fuel Cells (SOFCs) owing to their good ionic transport properties as well as adequate mechanical and chemical stabilities.[1] In particular, Y-doped $BaZrO_3$ has been widely studied as this material shows very high protonic conductivities of up to $1\times10^{-2}$ S cm$^{-1}$ at 450°C when exposed to water vapor.[2,3] When $BaZrO_3$ is doped with yttrium, the charge imbalance between the yttrium and zirconium ions is accommodated by the generation of oxygen vacancies leading to the composition $BaZr_{1-x}Y_xO_{3-x}$. These oxygen vacancies are responsible for the moderate oxygen-anion conductivity at high temperatures.[3] Under wet conditions, protons can be incorporated in the material through absorption of water molecules that dissociate forming hydroxide ions and filling the oxygen vacancies. The material can be fully hydrated, represented by the chemical formula $BaZr_{1-x}Y_xO_{3-x}(OH)_x$, or partially hydrated represented more generally as $BaZr_{1-x}Y_xO_{3-x/2+y/2}(OH)_y$ (y < x) with oxygen vacancies remaining. The high ionic conductivities achieved in wet conditions are promising for applications of this material in intermediate-temperature SOFCs.[3]

Y-substituted $BaZrO_3$ can be prepared with high concentrations of yttrium (up to 60%[4]) and a number of studies[4,5] have reported a dependence of the ionic conductivity on the yttrium content. As a first guess, one would expect that the ionic conductivity increases with the doping concentration following the increase in the number of charge carriers. However, it was shown that an optimum conductivity is observed for a dopant concentration of around 10%-20%.[4,5] In fact, many experimental and theoretical observations suggest that the protons are trapped close to the yttrium dopant sites increasing the activation barrier for diffusion and reducing the overall ionic conductivity.[6–8] This proposed proton trapping has attracted considerable attention in recent years both from an experimental and from a theoretical point of view.[8–17]

In hydrated Y-doped $BaZrO_3$, proton conduction occurs through series of transfers (between neighboring oxygens) and rotations (around an oxygen). Static density functional theory calculations (DFT) and molecular dynamics simulations (based on DFT or reactive force fields) have been proposed as powerful methods to estimate energy barriers for the intra- and inter-octahedral transfers as well as for rotations. Static DFT calculations have shown that local distortions arise close to the dopant atom, usually leading to stronger hydrogen bonds, which could explain the trapping effect.[10–13] However, local distortions also induce shorter O-O distances, which would tend to decrease the activation energies for proton jumping. It is worth noting that trapping does not necessarily imply that jumps do not happen but rather that protons spend a significant amount of time in the vicinity of the dopant and that the jumps are local rather than involving long-range transport. In many studies, local activation energies are combined with a jump-diffusion model to predict long-range energy barriers and diffusion coefficients. The resulting activation energies are usually in good agreement with experiments and trends observed for different dopant atoms could be reproduced by Björketun *et al.*[11] In this context, Blanc *et al.* have defined specific proton environments in the perovskite material depending on the position of the proton with respect to the dopant atom (see Figure 1)[14]. In their DFT calculations, the configurations with protons close to the yttrium atoms indeed correspond to the lowest energy structures.

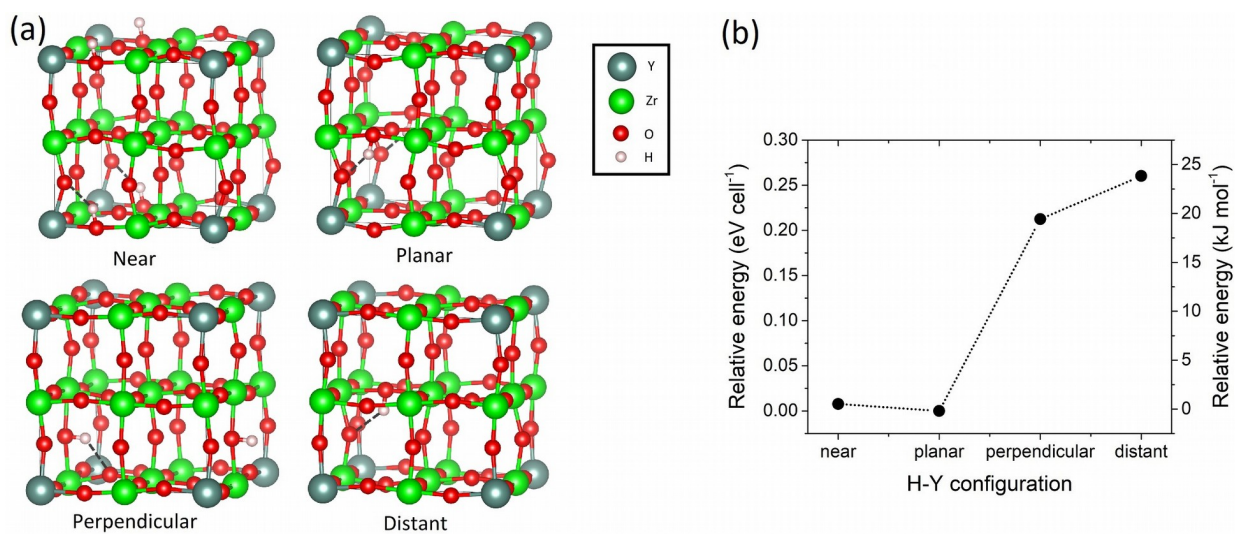

Figure 1. (a) Four possible proton-yttrium (H-Y) configurations in 2x2x2 supercells with a 12.5% dopant concentration as defined by Blanc *et al.*[14] Barium atoms in the A sites of the perovskite structure (at the centre of the Zr/Y-O cubes) are not shown for clarity. (b) Relative energies of these cells with respect to the lowest energy structure after geometry optimization, extracted from the data of Blanc *et al.*[14]

While static DFT calculations are very useful to identify proton diffusion pathways and the effect of specific local distortions, thermal fluctuations are always present at the temperatures used in the experiments (200°C - 700°C). These dynamic distortions may alter the picture obtained from a static point of view. DFT-MD simulations have the advantage of being able to capture both proton hopping and thermal fluctuations that occur on short time scales as demonstrated by many studies for water in the bulk liquid phase and at interfaces.[19–21] More recently, Fronzi *et al.* applied DFT-MD simulations on proton conduction in undoped $BaZrO_3$ and reported that proton diffusion is enhanced when compressive strain is applied on the structure.[22] The authors interpret this increase in diffusion to the shorter O-O distances observed in the compressed material. The effect of strain on proton diffusion has also been demonstrated in Y-doped $BaZrO_3$ for two dopant concentrations (1% and 12.5%) using MD simulations with a reactive force field based on an Empirical Valence Bond description.[23] In these dynamical studies and other related ones,[15–17] no strong trapping is however observed. Raiteri et al. suggest that for low dopant amounts (less than 15%), the trapping probability is low and thus not often observed, or hard to characterize, from MD simulations.[16] As such, there still remains a need to develop a clear understanding of the trapping effect and the effect of lattice dynamics on mobility at intermediate-to-high temperatures – the temperatures relevant to the operation of the device.

In this work, we combine static DFT calculations and *ab initio* molecular dynamics to obtain insights into proton diffusion and trapping in fully hydrated Y-doped $BaZrO_3$. Static DFT calculations are performed on structures with two different Y-dopant concentrations (12.5% and 25%) and are used to determine the energies of a large number of configurations and correlate energies with distortions of the lattice. The effect of the relative position of yttrium atoms is explored by using a variety of structures in both static and DFT-MD simulations; we assume that no oxygen vacancies are present to simplify the analysis. *Ab initio* molecular dynamics simulations are then conducted on structures with a 12.5% dopant concentration at three temperatures. From the generated trajectories, the proton diffusion was analyzed through mean square displacements as well as via the numbers of proton jumps between different environments. The results show a relationship between specific local geometries and the occurrence of proton jumps. In particular, intra-octahedral jumps are only seen when short hydrogen bond lengths are formed, more common at high temperatures. Activation energies calculated are in good agreement with previously reported data corresponding to local proton motion.

## 2. Computational methods

### 2.1 Static DFT calculations

For the static DFT study, results reported in the literature[12,14] for a dopant concentration of 12.5% were combined with results obtained in this work for a dopant concentration of 25%. All geometry optimizations were achieved using the PBE functional[24,25] and were done with a 2x2x2 supercell having the generic formula $BaZr_{1-x}Y_xO_{3-x}(OH)_x$. The supercells contain 41 atoms for the 12.5% dopant concentration and 42 atoms for the 25% dopant concentration, i.e., the material is considered fully protonated and there are no oxygen vacancies. Four of the structures with a 12.5% dopant concentration and all 26 structures with a 25% dopant concentration were optimized using CASTEP[26], a basis set cut-off of 680 eV, an energy convergence criterion of $10^{-6}$ eV (as in Ref.[14]) and a 3x3x3 Monkhorst-Pack[27] k-point mesh. The 96 other structures with a 12.5% dopant concentration have been reported in a previous work[12] and were generated, using VASP,[28,29] as follows. The positions of all atoms in an initial 2x2x2 supercell with an yttrium dopant, without any protons, are relaxed. To compensate the charge imbalance, VASP automatically uses a uniform positive background gas. A proton is then placed in all possible accommodating sites (24 oxygens x 4 possible orientations),[12] and its position is relaxed while keeping all the other atoms fixed. For these 96 proton binding structures, started from the lowest energy Glazer distortion, the geometry optimizations were done using the Generalized Gradient Approximation (GGA) PBE functional. A basis set cut-off of 600 eV, an electronic energy convergence criterion of $10^{-4}$ eV and a 2x2x2 Monkhorst–Pack k-point mesh were used. The geometry optimization energy convergence criterion was set to $10^{-3}$ eV. In the end, a total of 100 structures with a 12.5% dopant concentration and 26 structures with a 25% dopant concentration was obtained. VESTA[18] software was used for plotting all the structures shown in this article.

## 2.2 DFT-MD calculations

For the DFT-MD study, larger 4x4x4 supercells containing a total of 328 atoms (including 8 protons) and having a single dopant concentration of 12.5% were simulated. All the supercell. (b) YC: Some of the yttrium dopant atoms are nearest neighbors (clustered). (c) YP: Some of the yttrium dopant atoms are in the planar configuration. In all cases, there are 8 protons with 2 of them arranged in each H-Y configuration (near, planar, perpendicular, distant).

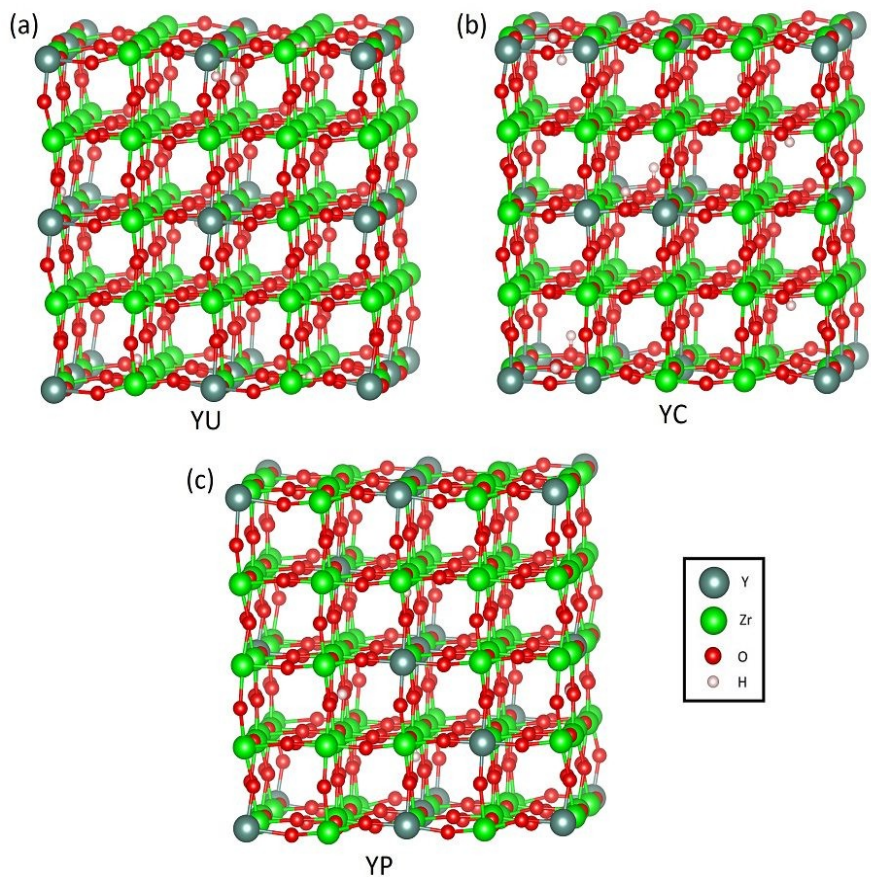

Figure 2. Starting configurations used for the DFT-MD simulations performed on cells containing 12.5% yttrium. (a) YU: Yttrium dopant atoms are distributed uniformly in the 4x4x4

calculations were done with the CP2K software[30] using the PBE functional and run in the NPT ensemble under atmospheric pressure. The cell is cubic with an initial lattice size of 17.04 Å. The Gaussian Plane Wave (GPW) implementation uses a Gaussian DZVP basis set with GTH pseudopotentials,[31–33] a 400 eV plane wave cut-off, with only the Gamma point, and a $6 \times 10^{-7}$ eV electronic energy convergence criteria. Three different distributions of the yttrium dopant atoms were considered: a uniform distribution (noted YU), a clustered configuration where two pairs of yttrium atoms are nearest neighbors (noted YC) and a planar configuration where two pairs of yttrium atoms are next nearest neighbors (noted YP). This is illustrated in Figure 2, which shows the initial configurations for the DFT-MD simulations. To obtain the starting configurations, 8 protons were distributed in the system so that there are 2 protons in each H-Y configuration as defined in Figure 1 (near, planar, perpendicular, distant)[14]. Three temperatures (500 K, 750 K and 1000 K) were investigated in order to estimate the activation energies related to proton diffusion. All the DFT-MD simulations were run for 20 ps with a 0.5 fs timestep.

**3. Results and discussion**

**3.1 Static DFT calculations**

As discussed in the introduction, the relative positions of the protons and yttrium atoms, i.e., the H-Y configuration, have a large influence on the energy of the 12.5% dopant concentration cells. To investigate the impact of the relative positions of the yttrium atoms, i.e., the Y-Y configuration, the configurational energies and cell geometries of 26 structures with a dopant concentration of 25% are first analyzed. Figure 3 shows the relative energies of these structures with respect to the lowest energy structure and illustrates the Y-Y configurations.

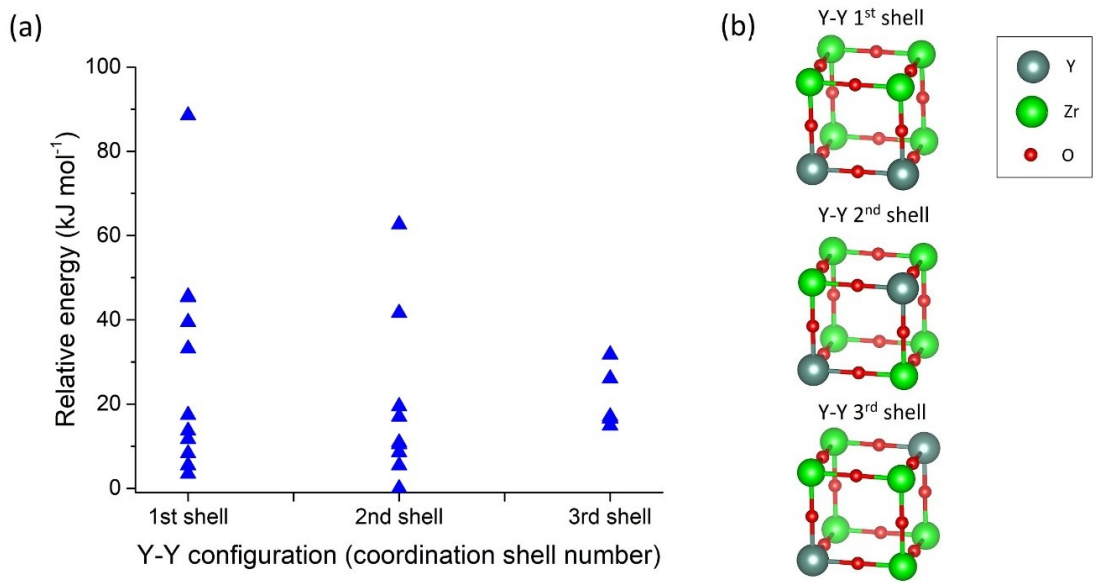

Figure 3. (a) Relative configurational energies with respect to the lowest energy structure for 26 cells having a dopant concentration of 25% and different Y-Y configurations. (b) Schematic representation of the different Y-Y coordination shells (these structures are not optimised and simply illustrate the relative positions of the yttrium atoms).

From Figure 3, it is clear that a larger range of energies is observed for 1$^{st}$ and 2$^{nd}$ shell than for 3$^{rd}$ shell configurations. We also observe that the lowest energies are obtained for protons nearby Y-Y in 1$^{st}$ and 2$^{nd}$ shell configurations. We note, however, that the Y-Y arrangements are determined during the high temperature synthesis of the material to form $BaZr_{1-x}Y_xO_{3-x}$: hydration is unlikely to change the Y-Y configurations at least at moderate temperatures. Due to the high temperatures used to prepare these materials, yttrium atoms are likely to be randomly distributed in the lattice as suggested by previous researchers.[1,34] Different syntheses routes may lead to different materials properties which could perhaps be related to different dopant spatial distributions.[35,36]

To characterize the lattice distortions due to the presence of an yttrium atom, the metal-oxygen (Zr-O, Y-O) and metal-metal (Zr-B, Y-B) interatomic distances for different dopant concentrations (B = Zr, Y) are examined. Figures 4.a and 4.b show a comparison of the average interatomic distances calculated from the static DFT calculations and the experimental values[32]. The metal-oxygen and metal-metal distances obtained from DFT are in good agreement with experiments although slightly overestimated (by around 1-5%). These results are consistent with other simulation results published on similar materials[37] and show that the PBE functional is suitable for such a study. From Figures 4.a and 4.b, it is clear that while the average metal-oxygen distance is larger for yttrium-oxygen than for zirconium-oxygen, the average metal-metal interatomic distances are relatively unaffected by the presence of yttrium. This suggests that there is a bending of the B-O-B units, consistent with the octahedral tilting reported previously.[37,38] Figure 4.c shows the interatomic distances for different environments around the oxygen in the presence or absence of protons. While the presence of a single yttrium atom only leads to a limited variation in the metal-metal distance, the presence of two yttrium atoms results in a noticeable decrease in distance. We also note that the metal-metal distances for atoms surrounding an OH unit are systematically larger than in the absence of proton. This could be due to steric or electrostatic repulsions between the proton and the adjacent Zr/Y atoms.

To investigate the effect of local distortions close to the H and Y atoms on the energy of the cells, we focus on the 100 structures with a 12.5% dopant concentration that contain only one Y and proton per supercell. (In the 25% doped structures, there are two protons and two yttrium atoms, and it is thus difficult to separate the effects of the two H-Y configurations; furthermore the relative energies of the different Y-Y configurations would need to be accounted for).

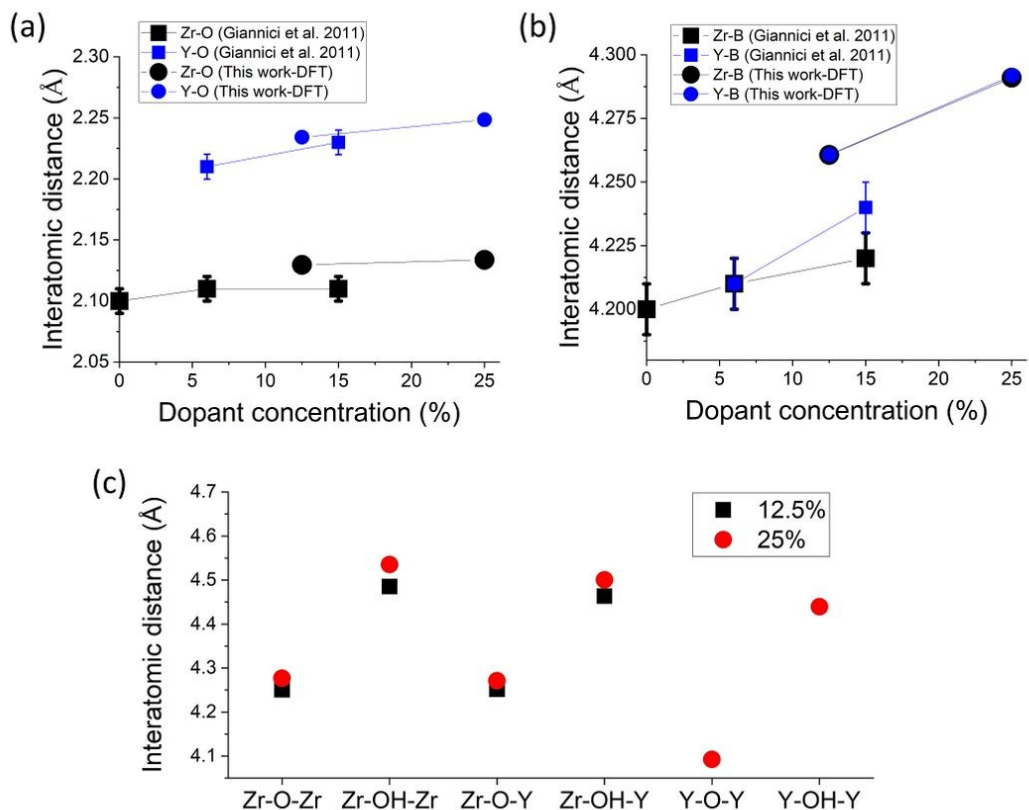

Figure 4. (a) and (b) Metal-oxygen and metal-metal interatomic distances as a function of the Y-dopant concentration calculated in this work from static DFT (12.5 and 25%) and determined experimentally from X-ray diffraction and extended X-ray fine structure (EXAFS) analysis (0, 6 and 15%) by Giannici et al.[37] (c) Metal-metal interatomic distances for specific metal-oxygen-metal configurations obtained in this work.

To characterize the local distortions shown in Figure 5.a, we used a signed "B-OH-B bending distance" which is the non-zero component of the vector from the center of the solid black line connecting the two transition metals (Zr or Y) to the oxygen covalently bonded to the proton. The 100 structures are classified into two categories depending on the sign of the B-OH-B bending distance. If the B-OH-B bending distance is negative, the structure is assigned as "bent outwards" since the hydroxyl is outside of the interior BOB angle, the structure is otherwise assigned as "bent inwards" when the hydroxyl is on the inside of the interior BOB angle.

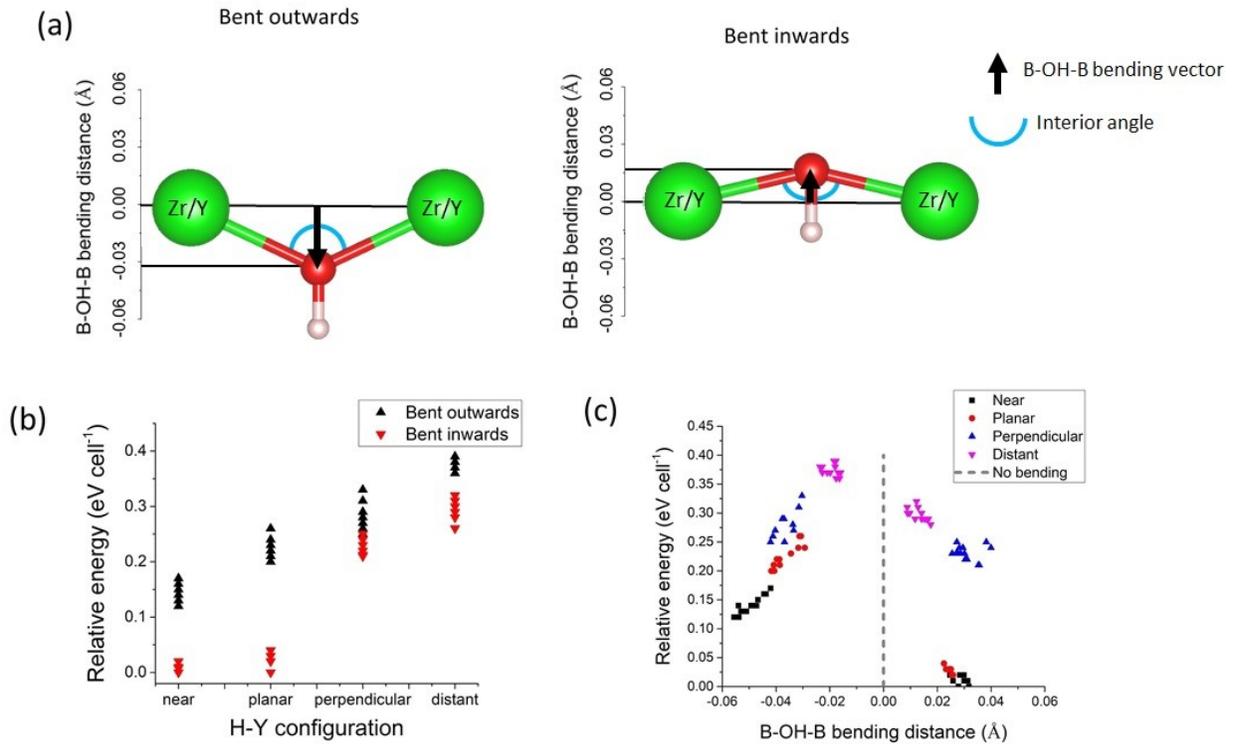

Figure 5. (a) Illustration of inwards and outwards bending, showing how the bending distance is quantified, a B-OH-B bending distance of 0 corresponding to colinear bonds (i.e., no bend). (b) Relative energies of the 100 structures with a 12.5% dopant concentration classified following their H-Y configuration and B-OH-B bending type. (c) Relative energies as a function of the B-OH-B bending distance.

Figure 5.b shows a plot of the relative cell energies as a function of the H-Y configuration. The different structures are colored according to their "bent inwards" or "bent outwards" bending status. It is very clear from this figure that, for a given H-Y configuration, all the structures which are "bent inwards" have lower energies than the "bent outwards" configurations and are thus more favorable. To confirm this assumption, we plot the relative cell energies as a function of the B-OH-B bending distance. This is shown in Figure 5.c where the different structures are now colored according to their H-Y configuration. While the trend is not perfect and while the energies of some structures overlap, there is a clear correlation between the cell energy and the type (inwards/outwards) and amplitude of the bending. Independent of the bending type, larger absolute distances (i.e., larger bendings) are correlated with lower energies. We note that all the points corresponding to specific H-Y configurations are grouped and that the lowest energies are observed for combined near/planar and bent inwards configurations. As far as we know this correlation between specific distortions and cell energies has not been highlighted before and confirms that the presence of yttrium dopant close to the protons can lead to low energy configurations not accessible otherwise.

We now investigate what features makes the lowest energy structures more stable. While, the proximity of protons, oxygen and yttrium atoms can lead to complex hydrogen bonding situations, which can influence both the local structure and proton dynamics,[12,39,40] here we focus on the characterization of these hydrogen bonds to determine how important this is in determining the calculated configurational energies. To describe the different environments, a number of angles and distances, all defined in Figure 6.a, are used. $dHO_1$ and $dHO_2$ are the distances between the proton and either the second or third nearest oxygen, respectively. $dHO_1$ is thus the traditional hydrogen bond length while dHO is the length of the covalent bond (distance to the nearest oxygen). $OHO_1$ corresponds to the hydrogen bond angle while $OBO_1$ and $OBO_2$ characterize the distortions around the metal atoms. Figures 6.b, 6.c, and 6d clearly show that the lowest energy configurations correspond to specific local environments where the hydrogen bonds are the strongest. Indeed, Figure 6.b shows that all the lowest energy structures correspond to hydrogen bond lengths smaller than 2.15Å and angles larger than 130°. Figure 6.c illustrates the fact that all the lowest energy structures correspond to situations where one of the hydrogen bond ($HO_1$) is noticeably stronger than the other one ($HO_2$). In these low energy environments, $dHO_1$ is clearly smaller than $dHO_2$. Interestingly, these stronger hydrogen bonds are correlated with less distorted $OBO_1$ angles, i.e., with angles close to a value of 90 degrees which would be seen in an ideal cubic structure (see Figure 6.d). It is worth noting that, for the remaining configurations, a number of data points are close to the "y = x" line. In these cases, the proton is engaged in two very similar hydrogen bonds. In the case of inwards bending, the formation of hydrogen bonds is intuitively favorable as it is consistent with the octahedral tilting while in the case of outwards bending, the formation of hydrogen bonds goes against the expected tilting. This phenomenon is illustrated in Figure S1.

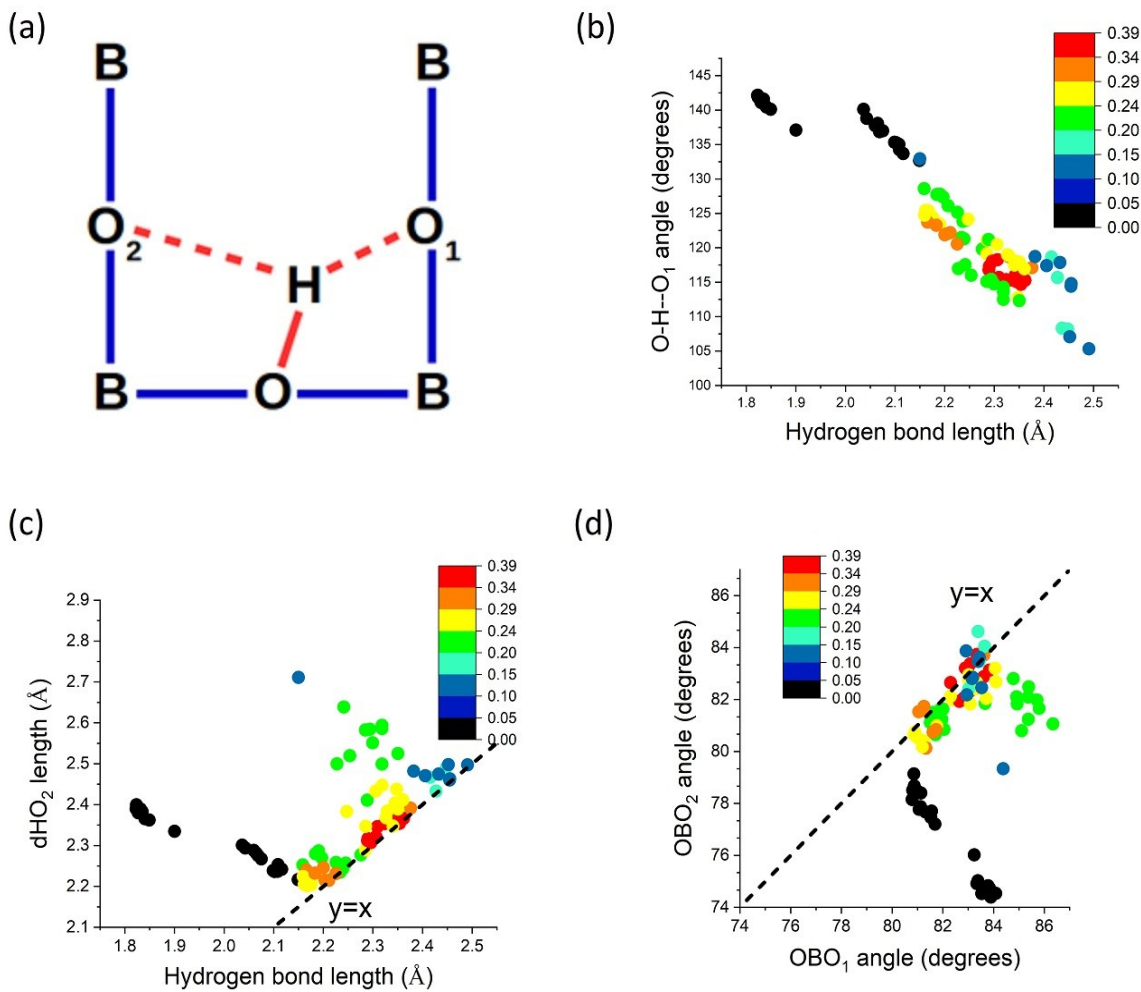

Figure 6. (a) Schematic representation of the local geometry around a proton and labels adopted to assess hydrogen bonding related geometries. (b) Hydrogen bond angle as a function of hydrogen bond length (i.e., $dHO_1$). (c) Distance between the proton and the 3$^{rd}$ nearest oxygen (i.e., $dHO_2$) as a function of the hydrogen bond length. (d) Metal-oxygen-metal angles around the hydrogen bond. For (b), (c), (d) the structures are colored according to their relative energy with respect to the lowest energy structure in black (energies are in eV cell$^{-1}$).

Overall, the static DFT calculations indicate that, in agreement with previous studies, the presence of yttrium as a dopant can lead to the existence of low energy configurations which could act as proton traps. These low energy structures are apparently stabilized by a strong hydrogen bond in a locally distorted environment. We now turn to DFT-MD simulations to investigate the existence and fate of such configurations in a dynamic system.

**3.2 DFT-MD calculations**

In order to obtain insight into the proton dynamics in Y-doped $BaZrO_3$ materials, a number of DFT-MD simulations were performed on 4x4x4 supercells having a single dopant concentration of 12.5% at three temperatures (500 K, 750 K, and 1000 K). In the smaller 2x2x2 supercells used above for the DFT analysis of the static systems, there was only one Y-Y arrangement, the different configurations arising from the different positions of the proton in the cell (Figure 1), however, the larger 4x4x4 cells require that we now consider the relative Y positions (as was performed for the smaller cells containing 25% Y, Figure 3). Three distributions of yttrium atoms were investigated (Figure 2). In one of the systems, the Y atoms are distributed uniformly and quite far apart from each other (YU); while in the clustered system (YC), some of the Y atoms are nearest neighbors; and in the last configuration, some of the Y atoms are next nearest neighbors, i.e., in the planar configuration (YP). The initial structures are generated so that the 8 protons are distributed equally in all H-Y configurations (2 protons in near, 2 in planar, 2 in perpendicular and 2 in distant). The molecular dynamics simulations are then conducted and the protons trajectories are recorded.

Figure 7.a provides an illustration of the environments visited by one of the protons along a trajectory for the YU system at 1000 K. While the proton is originally in a planar configuration at t = 0, it jumps between planar and near configurations ultimately visiting all possible H-Y configurations along the 20 ps trajectory. The actual trajectories of the 8 protons present in the same YU system at 1000 K are shown in Figure 7.b. The total residence time fractions of the protons in the different H-Y configurations, i.e., the total time protons spend in a given configuration divided by the total simulation time, for all systems are shown in Figure S2 (these time fractions are normalized by the available sites in each system in Figure S3). They confirm that all H-Y configurations are explored by protons. While the perpendicular and distant configurations are generally slightly less favorable than the near and planar ones (Figure S2), the differences between residences times between different sites are not as large as might be expected based on the static DFT calculations (Figure 1). However, direct comparison with predictions from static DFT energies is out of reach since these total residence time fractions are not fully converged for the 20 ps trajectories conducted here (Figure S4). Note that the initial configurations of the protons were generated to explore as many microstates as possible and since all of the configurations are sampled on the timescales used here, the short times employed in the MD runs should still allow comparisons between structural properties obtained in the static calculations and by DFT-MD.

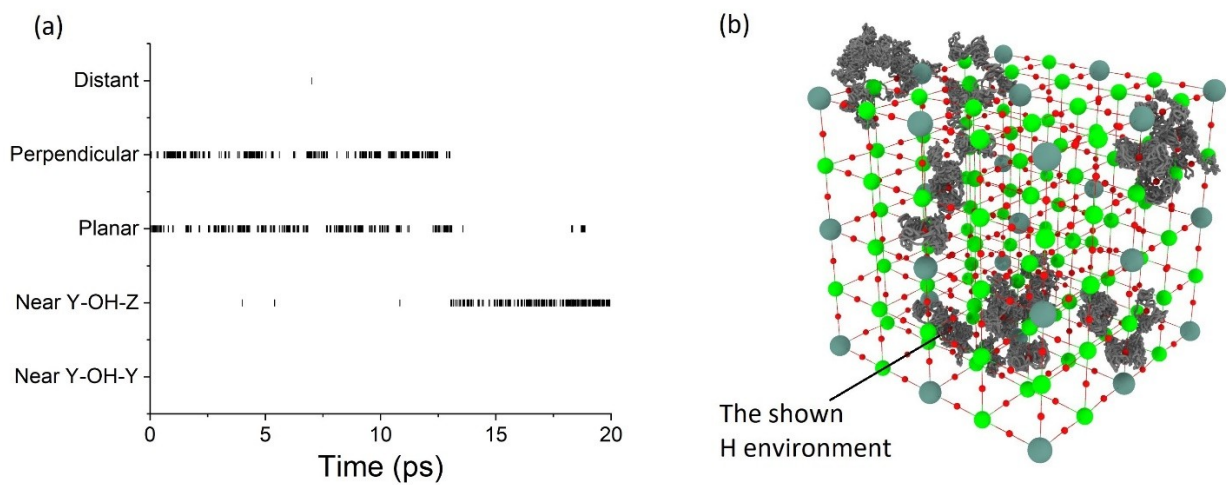

Figure 7. (a) Illustration of the various environments visited by a single proton along a 20 ps trajectory for the YU system simulated at 1000 K. The near configuration is split into two environments depending on the proton having 1 or 2 yttrium atoms as nearest neighbors ("Near Y-OH-Z" and "Near Y-OH-Y"). Note that in YU and YP, there are no Y-OH-Y configurations so this environment cannot be visited in the trajectory shown here. (b) DFT-MD trajectories of the 8 protons present, visualized using the OVITO program[41]. The proton whose trajectory is analysed in (a) is indicated.

Since it is possible to distinguish the different proton environments with respect to the yttrium atoms (near, planar, perpendicular, distant), we can calculate various properties according to each H-Y configuration. Figure 8 shows the average residence time for each H-Y configuration, i.e., the average time that a proton stays in a given site before jumping to another site. For most of the configurations, the average residence time decreases as expected when the temperature increases. For the perpendicular configuration however, the average residence time stays short even at 500 K. In fact, the average residence time in the perpendicular configuration is the shortest for all the systems considered. In the static DFT calculations, the configurational energies for near and planar are low and close to each other, while the perpendicular and distant configurations have similar and high energies. Following these results, longer residence times could be expected for near and planar configurations, but this is not what is observed here. A low energy barrier for proton transfers between near and planar might explain the relatively small residence times determined for these H-Y configurations. In addition, fluctuations of the oxygen-oxygen distances are probably an important component in defining the time between jumps, as discussed below.

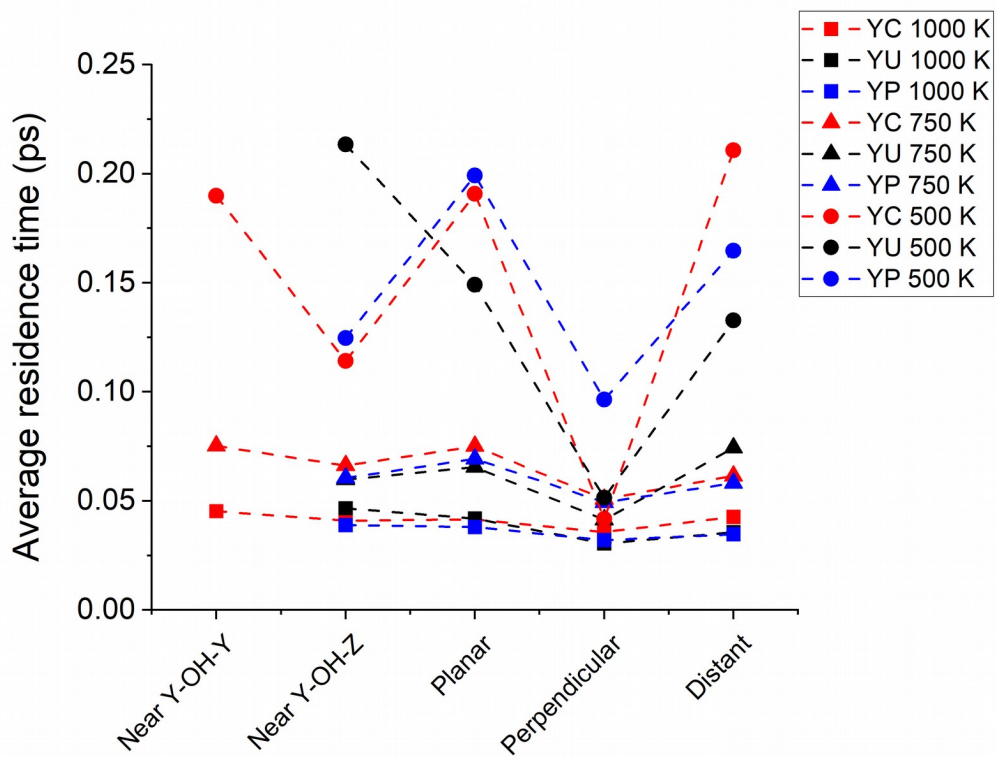

Figure 8. Average residence time of a proton in the various H-Y configurations.

Proton transfers and rotations between environments are examined by counting how many proton jumps occur from a given type of site to another. The total number of rotations and intra-octahedral transfers are given in Table S1. It is worth noting that no inter-octahedral jump is observed in any of the simulated systems. This is consistent with an activation barrier being higher for this event compared to rotations and intra-octahedral jumps.[10,38,42] There are approximately 20 times more rotations than intra-octahedral transfers which is consistent with activation barriers reported previously.[10] When two yttrium atoms are nearest neighbors (YC), there are much more near to near proton jumps. This could be because the local distortions resulting from the proximity of two yttrium atoms reduces the oxygen-oxygen distances thereby facilitating proton rotations and jumps as discussed in Ref.[10,38] In addition, most of the jumps are gathered on the diagonals, meaning that they do not change the type of H-Y configuration. The exceptions are the rotations between perpendicular and planar and the intra-octahedral jumps between planar and near.

To relate these jump events to local distortions, the jumps are now divided into jumps from inwards and outwards configurations. The number of rotations classified in this way are given in Table S2 and plotted in Figure 9 along with the fraction of inwards and outwards states (note that while the trajectories are short, these fractions are quite well converged as can be seen in Figure S5). This analysis shows that all intra-octahedral transfers occur when the protons are in the outwards configuration. Figure 9 also shows that, in most cases, even though on average there are more protons in the outwards states, more rotations originate from the bent inwards configuration. This indicates that inwards configurations are more favorable for rotations. These considerations, combined with the lowest energies observed for near and planar H-Y configurations with inwards bending indicate that the presence of yttrium atoms favor inwards bending, and thus rotations, thereby reducing the occurrence of intra-octahedral jumps and thus long-range diffusion. This could, at least partly, explain the proton trapping observed in Y-doped hydrated $BaZrO_3$. It is worth nothing that while there are fewer protons in inwards configurations than in outwards configurations at the temperatures simulated here, this proportion decreases with increasing temperature, in agreement with a lower energy for inwards configurations. A rough estimate of the energy difference between inwards and outwards from their populations at 500 K and 1000 K indicates that the inwards configurations energies are around 8-10 kJ mol$^{-1}$ lower than the outwards (corresponding to 0.10 eV cell$^{-1}$ in units comparable to the static DFT calculations reported in this work) which is consistent with the range of values observed in Figure 5.c.

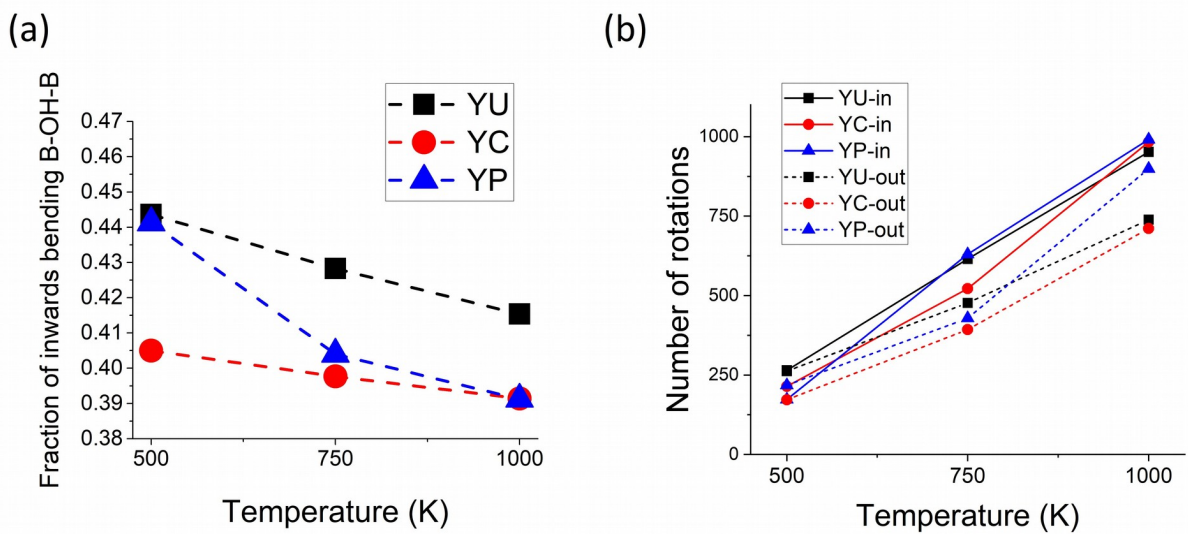

Figure 9: (a) Average fraction of inwards bending states observed in the different systems. (b) Total number of rotations originating from inwards/outwards configurations.

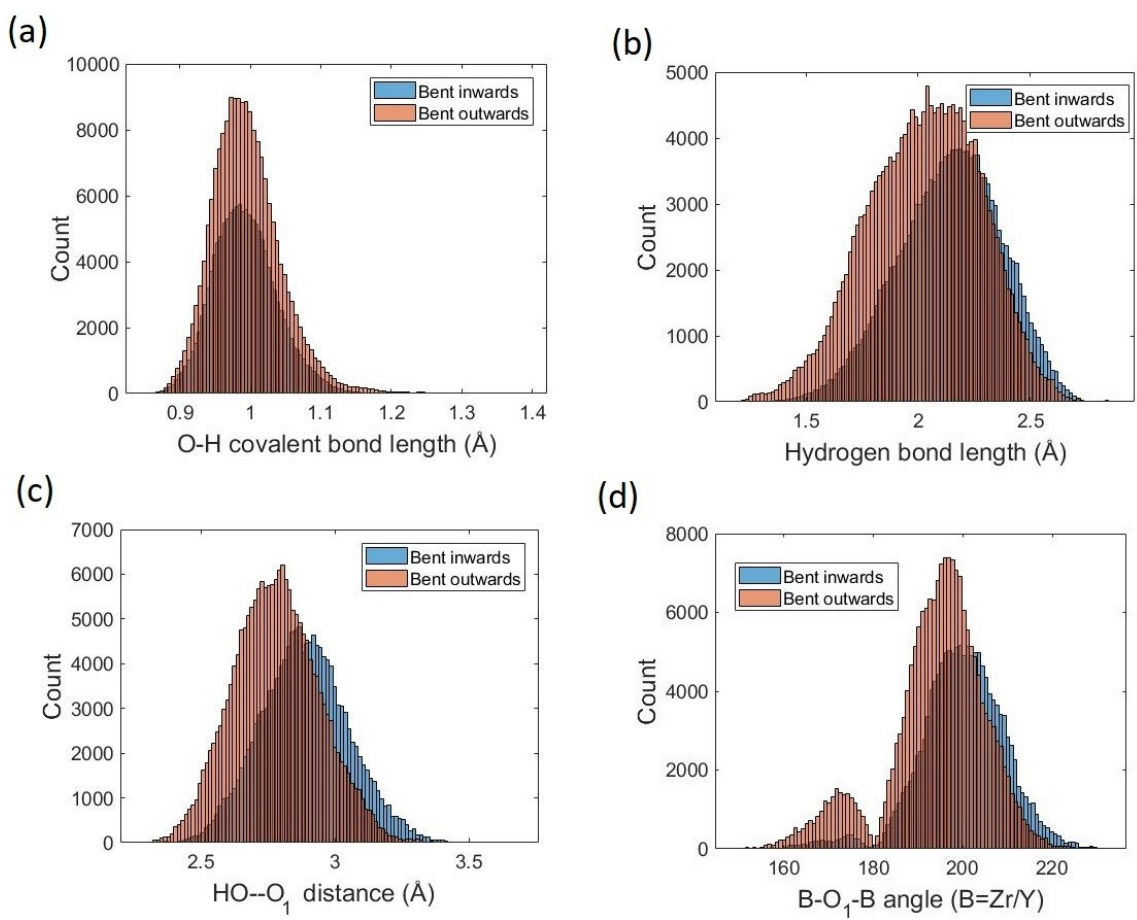

Figure 10. Several distances for different proton positions analyzed according to inwards/outwards bending states in the YU system at 750 K: (a) distance between the proton and its closest oxygen atom, (b) hydrogen bond length, (c) distance between two oxygen atoms forming a covalent and a hydrogen bond to the same proton, (d) bending of B-O$_1$-B angles.

We now try to correlate these results with a description of the hydrogen bonding in the material. To begin with, the same geometric descriptors as in Figure 6 are used to see if major differences between inwards and outwards configurations can be identified. The plots obtained are shown in Figure S6. As expected, due to thermal motion, the clouds of points span a much wider area than in the case of the static DFT calculations. Moreover, the hydrogen bond lengths and the $OBO_1$ and $OBO_2$ angles are on average a bit larger for the inwards configurations. To obtain further insights into the difference between inwards and outwards configurations, histograms of several descriptors of the hydrogen bonding environment are plotted; the histograms were all checked to ensure that they are well converged over the simulated trajectory for all systems. The histograms for the YU system simulated at 750 K are given in Figure 10. Similar results are observed for the YC and YP systems at 750 K (Figure S7 and S8). Figure 10 shows that the covalent bond lengths are similar for inwards and outwards configurations and as such cannot be used to interpret the proton jumps results. In contrast, inwards and outwards configurations show some difference with regards to the hydrogen bond length and HO--$O_1$ distance. On average the outwards configurations show shorter hydrogen bond lengths – in contrast to what is observed in the static calculations – and shorter HO--$O_1$ distances which would tend to reduce the activation barriers for the jump to the H-bonded $O_1$ site and thus increase the jump rate. As such, intra-octahedral transfers are possible in the case of protons in outwards configurations but only rotations, in which the covalent OH bond is not broken, are observed for protons in the inwards configurations. It is worth highlighting here that local distortions, motion of oxygen that is covalently bonded to the proton, and rotation of proton can lead to inwards/outwards exchange and so the conclusions above do not imply that protons which are in inwards configurations are blocked for the entire trajectory.

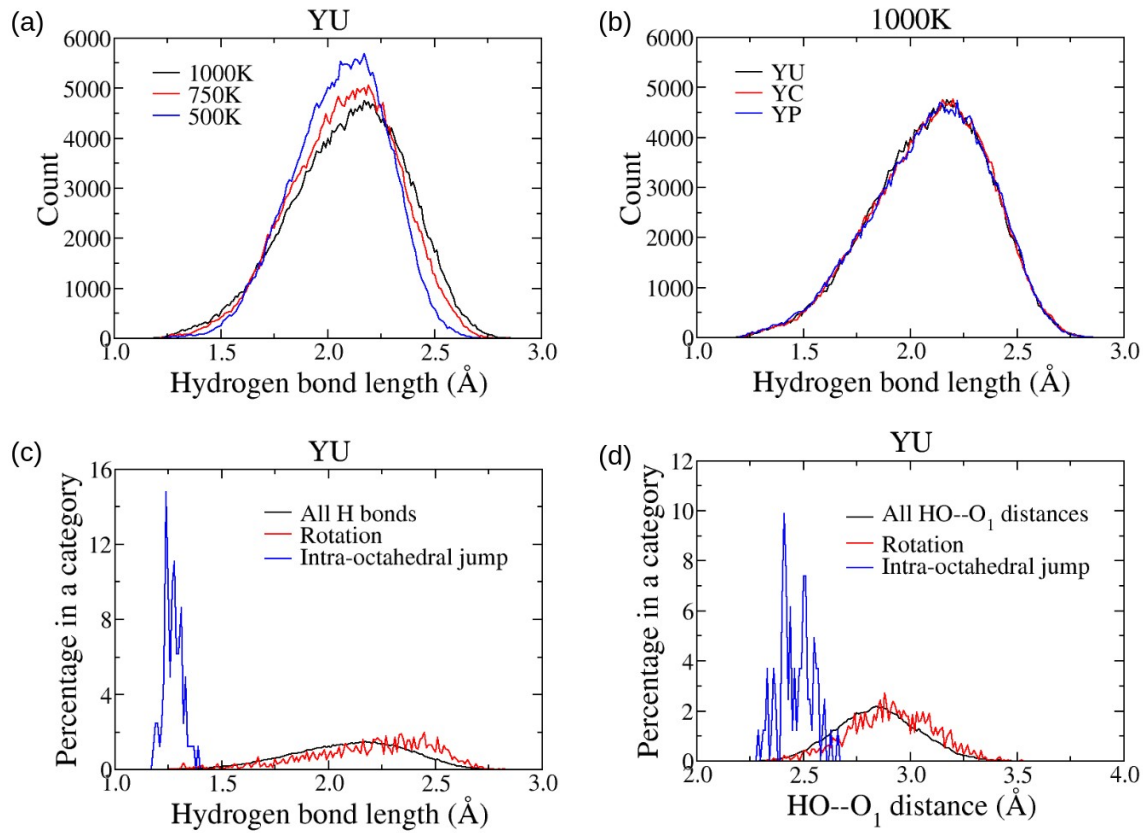

Figure 11. (a) Effect of temperature on the hydrogen bond lengths for the YU system. (b) Comparison of the hydrogen bond lengths for all the simulated systems at 1000 K. (c) and (d) Histograms of hydrogen bond lengths and HO--$O_1$ distances for the YU system at 1000 K. Black lines correspond to all the hydrogen bonds existing in the simulation box over time. Red and blue lines correspond to lengths at the time of a rotation or intra-octahedral jump, respectively. The number of distances in each category is very different (320,000 for the entire set of lengths, up to a few thousands for rotations and up to a few tens for intra-octahedral jumps) so, to facilitate the comparison, histograms are given as a percentage in a given category.

As proton jumps are highly correlated with hydrogen bonds in the structure, it is interesting to look more closely at the effect of temperature on the hydrogen bond length. Figure 11.a shows the histograms for all the hydrogen bond lengths existing in the YU system at the three temperatures studied here. Similar plots for YC and YP are given in Figure S9 and show once again that the Y-Y distribution does not affect significantly the results (as confirmed by the direct comparison of histograms in Figure 11.b). At higher temperatures, wider distributions of lengths are found, corresponding to larger fluctuations in the hydrogen bond lengths. Both shorter and larger lengths are explored, the former favoring intra-octahedral jumps, which occur preferentially when the hydrogen bonds, and HO--$O_1$ distances, are the shortest. To confirm this last statement, histograms of hydrogen bond lengths and HO--$O_1$ distances when jumps occur were plotted. Such histograms are shown in Figures 11.c and 11.d for the YU system and in Figure S9 for YC and YP (other temperatures give similar results). Here, histograms are normalised by the total number of distances found in each category (total, rotations, intra-octahedral jumps) to ease the comparison. From these histograms, it is very clear that intra-octahedral jumps only occur for short hydrogen bonds, a situation that happens more often when the temperature increases. For rotations, the maximum of the hydrogen bond length distribution is only slightly shifted to larger distances compared to the average bond length. The results clearly indicate that low temperature static calculations do not necessarily capture relevant motional processes at higher temperatures. It is worth pointing out that while the possible existence of oxygen vacancies is omitted in this study, such vacancies could be present, especially at high temperatures, and introduce compositional dependent chemical expansion effects which would potentially impact the proton dynamics.[43,44]

We now examine the proton diffusion in a more global fashion to see if the local dynamics observed lead to notable difference on long-range diffusion. From the DFT-MD trajectories, the proton diffusion coefficients are determined using the Einstein relation:

$$MSD = \langle (r_i^t - r_i^0)^2 \rangle = 6Dt$$

where the MSD is the mean square displacements, $r_i^t$ is the position of proton $i$ at time $t$, D is the diffusion coefficient and <> denotes an average over all protons. The MSDs calculated for the different trajectories are given in Figure S10 and the diffusion coefficients obtained from the MSDs and the corresponding activation energies, calculated using Arrhenius law and the values at different temperatures, are gathered in Table 1 along with values from the literature obtained experimentally or from classical molecular dynamics. Our calculations seem to overestimate the diffusion coefficients compared to previously reported values at similar temperatures. It is important to note that due to the computational limitations of DFT-MD, these simulations are done on a small number of protons and the timescales and system sizes may be too short to assess diffusion coefficients, which could explain this overestimation, as discussed in more detail below.

The activation energies calculated (20.4, 16.0 and 21.3 kJ mol$^{-1}$) are of the same order of magnitude for the three Y distributions studied. This is consistent with the relatively similar properties observed for these systems so far and with the fact that all structures contain a similar distribution of H-Y configurations (Figure S3) so that the global motion is always an average over protons in different configurations. The values calculated here are in the same range as the activation energy of 24.1 kJ mol$^{-1}$ reported in the computational work of Gomez *et al.* in which no yttrium is present and thus no trapping can occur.[38] Kinetic Monte Carlo (KMC) simulations, which simulated proton dynamics using 96 binding sites with a 12.5% dopant concentration, have predicted activation energies of 38 kJ mol$^{-1}$ for a single proton[45] and 43 kJ mol$^{-1}$ when 8 protons were included in the simulation.[46] In these simulations the protons spend a significant amount of time in Y-OH-Z sites but travel between these sites through perpendicular and planar sites. These activation energies are closer to the values of 40.5 kJ mol$^{-1}$ and 45.0 kJ mol$^{-1}$ determined experimentally.[1,8] KMC simulations are more suitable to characterize long-range diffusion because of their ability to sample rare events more efficiently. It is worth noting though that in both KMC and DFT-MD simulations, the distant sites were sampled only occasionally. In addition, the KMC and the classical molecular simulations considered in table 1[16,17] simulate proton motion on a potential energy surface derived from static DFT calculations, typically through geometry optimisations and Nudge Elastic Band calculations. This could also explain partly the discrepancies described here.

| System | Diffusion @500K | Diffusion @750K | Diffusion @1000K | Activation energies (kJ mol$^{-1}$) |
|---|---|---|---|---|
| YU | 2.60 | 6.51 | 35.96 | 20.4 |
| YC | 3.79 | 24.10 | 22.67 | 16.0 |
| YP | 2.19 | 16.26 | 26.29 | 21.3 |
| Ref [1] | 0.11 | 3.19 | 16.81 | 41.3 |
| Ref [8] | 0.07 | 1.16 | 3.30 | 16 (> approx. 1000K) <br> 45 (< 1000K) <br> 13.0 (T$_1$ measurements) |
| Ref [16] | 0.05 | 1.31 | 6.63 | 40.5 |
| Ref [17] | 0.01 | 0.26 | 1.49 | 43.4 |

Table 1. Diffusion coefficients (/ 10$^{-10}$ m$^2$ s$^{-1}$) and activation energies calculated for the 9 systems studied in this work having a dopant concentration of 12.5% and from the literature

corresponding to experimental (dopant concentrations of 10% and 20%)[1,8] and simulated values (dopant concentrations of 0.46% and 12.5%)[16,17].

Figure 7 shows that in the DFT-MD simulations, in some cases, a proton can jump between perpendicular and planar sites at the beginning of a trajectory before spending a longer time in Y-OH-Z sites with occasional excursions through planar sites. While not all protons show a similar trajectory and there are also occurrences of protons escaping from near or planar to distant sites, such a behavior, consistent with KMC simulations, has been observed for several protons in the DFT-MD trajectories. This is again a hint that the current DFT-MD trajectories may be too short to look at long range diffusion but that trapping could be observed more clearly in longer simulations, unfortunately hard to achieve due to computational cost.

Looking more closely at the experimental data, the calculated activation energies are intermediate between the activation energies determined at high (16.0 kJ mol$^{-1}$) and low temperatures (45.0 kJ mol$^{-1}$) by Yamazaki *et al*.[8] using impedance spectroscopy and H/D exchange experiments. The difference between the high and low temperature impedance data of 29 kJ mol$^{-1}$ was then ascribed to an association energy the protons need to overcome in order to experience long-range diffusion. In the same work, the authors also characterized proton motion by $T_1$ relaxation measurements using NMR experiments and reported a single activation energy of 13 kJ mol$^{-1}$ across the whole temperature range monitored (room temperature to 873 K). The $T_1$ measurements are a probe of local motion while the impedance spectroscopy gives information on long-range diffusion. This picture is consistent with the fact that, due to short time and length scales, our simulations probe mainly the local motion and only rare events of "trapped" to "trap-free" positions are sampled, which are responsible for the long-range transport. As a consequence, while some trapping is captured in the simulations, the rare events corresponding to protons going from "trapped" to "trap-free" states are not sampled with sufficient frequency to affect the activation energy in a significant way.

## 4. Conclusion

In this work, a combination of static DFT calculations and DFT-MD was used to study the structural properties and proton dynamics in Y-doped $BaZrO_3$ with two dopant concentrations. The presence of yttrium leads to significant lattice distortions, which, as quantified by, for example, the calculated metal-metal and metal-oxygen interatomic distances, are in agreement with reported experimental results. Going beyond a simple description of the average interatomic distances, the local distortions of the lattice have been characterized. This allowed us to identify two types of configurations related to the bending of the metal-oxygen-metal angle. The nature of the two configurations, labelled inwards bending and outwards bending, were shown to affect the energies of the systems. In particular, i) the larger the bending, the lower the energy and ii) the lowest energies were observed for structures combining near and planar configurations with an inwards bending. The lowest energy structures were the ones with the strongest hydrogen bonds, which is only possible in the inwards configuration in the static calculations.

DFT-MD studies allowed us to explore the effect of temperature on the local structures and proton trajectories. A detailed analysis of the proton jumps according to the different H-Y configurations showed that inwards configurations favor rotations while intra-octahedral jumps are only observed from outwards configurations. In fact, in contrast to static DFT calculations, the DFT-MD simulations indicate that outwards bending leads to shorter hydrogen bond lengths and HO--$O_1$ distances on average. This is consistent with favoring intra-octahedral jumps as a decrease in the HO--$O_1$ distance should decrease the energy barrier for the jump. The DFT-MD simulations confirm very clearly that intra-octahedral jumps only occur when the hydrogen bond lengths and H--$O_1$ distances are the shortest. It is important to note that local distortions and rotations of protons can lead to inwards/outwards exchange. As a consequence, the statements above do not imply that protons in inwards configurations at a given time are blocked for the entire trajectory. The proton diffusion coefficients determined at three different temperatures allowed for the extraction of activation energies, giving values in the same range as some experimental results that probe short range motion but in general lower than previously reported simulation results using KMC or classical molecular simulations based on potential energy

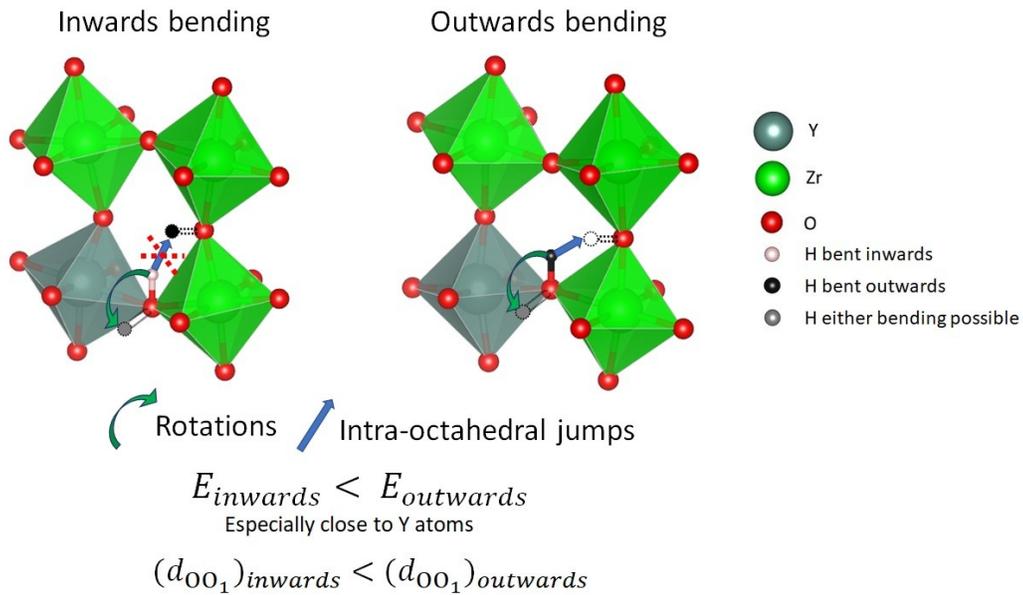

Figure 12. Scheme summarizing the main conclusions of this work regarding proton transfers. Static DFT calculations have permitted the identification of two configurations, namely inwards and outwards bending, which favor different proton transfers. The inwards bending is usually lower in energy, especially close to dopant atoms, and favors rotations with no intra-octahedral

jumps observed in the DFT-MD simulations conducted here. The outwards bending on the contrary allows for intra-octahedral proton transfers although still less frequent than rotations. These observations are consistent with proton trapping close to the yttrium dopant.

surfaces determined using static DFT methods. While trapping is believed to happen, it only affects the long-range diffusion and the local motion of protons is largely unaffected. As a consequence, the simulations reported in this work, probably too short to study long-range diffusion, do not capture the larger activation energies for long range proton diffusion. Overall, our results demonstrate the importance of the bending for proton dynamics and as local distortions are related with the yttrium content, this work is a new step towards understanding the existence of an optimal dopant concentration for the proton conduction. The results highlight the role of temperature and lattice motion on controlling the mechanisms for proton motion.

## ASSOCIATED CONTENT

**Supporting Information**. An illustration of the octahedral tilting, total residence time fractions, number of H-Y configuration sites in the systems, numbers of proton jumps, fractions of inwards and outwards bending as a function of time, characterizations of the interatomic distances related to the hydrogen bonding, mean square displacements.

## AUTHOR INFORMATION

The authors declare no competing financial interests.

## ACKNOWLEDGMENT

We thank Dr Bartomeu Monserrat, Dr Rachel Kerber and Dr Frédéric Blanc for fruitful discussions. C. M. acknowledges an Oppenheimer Research Fellowship from the School of Physical Sciences from the University of Cambridge. This project has received funding from the European Research Council (ERC) under the European Union's Horizon 2020 research and innovation programme (grant agreement no. 714581). Via our membership of the UK's HEC Materials Chemistry Consortium, which is funded by EPSRC (EP/L000202), this work used the ARCHER UK National Supercomputing Service. MAG' work was supported by the National Science Foundation under grant DMR-1709975 and the Mount Holyoke College Department of Chemistry. Computational resources were provided in part by the MERCURY consortium under NSF grant CHE-1626238. Data access statement: original data supporting this publication are available as Supporting Information.